\documentclass[aps, preprint, nofootinbib,preprintnumbers,eqsecnum,superscriptaddress,sort]{revtex4}
\pdfoutput=1
\usepackage{amssymb,amsmath,amsfonts,amsthm,latexsym,mathrsfs}
\usepackage{graphics, graphicx,color,epsfig,subfigure}
\usepackage[small,hang,justification=raggedright,singlelinecheck=on]{caption}
\usepackage[
      colorlinks=true,
      linkcolor=blue,
      urlcolor=blue,
      filecolor=black,
      citecolor=red,
      pdfstartview=FitV,
      pdftitle={},
        pdfauthor={Gary Horowitz, Jorge Santos},
        pdfsubject={},
        pdfkeywords={},
        pdfpagemode=None,
        bookmarksopen=true
      ]{hyperref}
\numberwithin{equation}{section}
\def \be {\begin{equation}}
\def \ee {\end{equation}}
\def \bea {\begin{eqnarray}}
\def \eea {\end{eqnarray}}
\def \dd {\mathrm{d}}

\newcommand{\bln}{\begin{align}}
\newcommand{\eln}{\end{align}}
\newcommand{\bst}{\begin{split}}
\newcommand{\est}{\end{split}}
\newcommand{\bi}{\begin{itemize}}
\newcommand{\ei}{\end{itemize}}
\newcommand{\ben}{\begin{enumerate}}
\newcommand{\een}{\end{enumerate}}

\def\eeq{\end{equation}}

\newcommand\smallO{
  \mathchoice
    {{\scriptstyle\mathcal{O}}}% \displaystyle
    {{\scriptstyle\mathcal{O}}}% \textstyle
    {{\scriptscriptstyle\mathcal{O}}}% \scriptstyle
    {\scalebox{.7}{$\scriptscriptstyle\mathcal{O}$}}%\scriptscriptstyle
  }

\begin{document}
%---Title and Abstract ------------------------------------------------
\title {Testing the Weak Gravity - Cosmic Censorship Connection}

\author{Toby Crisford}
\email{tc393@cam.ac.uk}
\affiliation{Department of Applied Mathematics and Theoretical Physics, University of Cambridge, Wilberforce Road, Cambridge CB3 0WA, UK}

\author{Gary T. Horowitz}
\email{horowitz@ucsb.edu}
\affiliation{Department of Physics, UCSB, Santa Barbara, CA 93106}

\author{Jorge E. Santos}
\email{jss55@cam.ac.uk}
\affiliation{Department of Applied Mathematics and Theoretical Physics, University of Cambridge, Wilberforce Road, Cambridge CB3 0WA, UK \vspace{1 cm}}

\begin{abstract}\noindent{
A surprising connection between the weak gravity conjecture and cosmic censorship has recently been proposed. In particular, it was argued that a  promising class of counterexamples to  cosmic censorship in four-dimensional Einstein-Maxwell-$\Lambda$ theory would be removed if charged particles (with sufficient charge) were present. We test this idea and find that indeed if the weak gravity conjecture is true, one cannot violate cosmic censorship this way. Remarkably, the minimum value of charge required to preserve cosmic censorship appears to agree precisely with that proposed by the weak gravity conjecture.
}
\end{abstract}
%\pacs{}
\maketitle

 %---Table of Contents ----------------------------------------------------
%\tableofcontents

%--- Manuscript ----------------------------------------------------
\section{Introduction and Summary}

The idea of cosmic censorship was proposed almost fifty years ago \cite{Penrose:1969pc}. By now there is ample  evidence that it does not hold in more than four dimensions. This is because there are unstable black holes in higher dimensions \cite{Gregory:1993vy, Hubeny:2002xn,Santos:2015iua},  and numerical evolution shows that the horizons pinch off in finite time \cite{Lehner:2010pn,Figueras:2015hkb, Figueras:2017zwa}.
This produces  regions of arbitrarily large curvature generically from smooth initial data.  Moreover, the mechanism behind all of these violations is essentially the same: the horizon develops a hierarchy of scales, and wants to break in a manner conjectured by Gregory and Laflamme  in \cite{Gregory:1993vy}. Recently, a  promising class of counterexamples to cosmic censorship in four dimensions with asymptotically anti-de Sitter (AdS) boundary conditions has been conjectured in \cite{Horowitz:2016ezu} with ample numerical evidence in its favour provided in \cite{Crisford:2017zpi}.  They are based on coupling gravity to a Maxwell field (and $\Lambda < 0$).
 
A seemingly unrelated idea is the weak gravity conjecture \cite{ArkaniHamed:2006dz} which was proposed about a decade ago. This states that any consistent theory of quantum gravity must have a stable particle with $q/m \ge 1$. It can be loosely interpreted as saying that gravity should always be the weakest force, since the gravitational attraction between two such particles is always less than the electrostatic repulsion. This also implies that extremal charged black holes are not stable, but can decay by Schwinger pair creation even though the Hawking temperature vanishes.

It has been suggested \cite{Vafa} that these conjectures might be related in the sense that assuming the weak gravity conjecture might remove the new counterexamples and preserve cosmic censorship. We will show that this is indeed correct.
 
Since weak gravity is a conjecture about quantum gravity and cosmic censorship is a conjecture about classical general relativity, to relate them we take a classical limit of the weak gravity conjecture. This means that we should include a charged scalar field in any proposed counterexample involving Maxwell 
fields\footnote{We will assume the weak gravity conjecture is satisfied by a charged scalar particle. If it is a fermion, one must do a more complicated analysis involving quantum fields in curved spacetime to investigate its effect on cosmic censorship.}.
  
As we review below, the counterexamples to cosmic censorship involve situations where an electric field grows in time without bound, producing arbitrarily large curvature that is visible to infinity. It is intuitively clear that if a charged scalar field is added with  large enough charge, the Einstein-Maxwell solution will become unstable to forming a nonzero scalar field. This is directly analogous to the instability of a charged black hole to form scalar ``hair" in a holographic superconductor \cite{Gubser:2008px, Hartnoll:2008vx}. It can be viewed as the classical analog of pair creating a cloud of charged particles.  If the solution becomes unstable, the previous counterexamples are no longer valid since it would require fine tuning to keep the scalar field zero. There remain two key questions: (1) If the scalar field is nonzero, can one still violate cosmic censorship?  (2) How does the minimum charge required for instability compare with the value predicted by the weak gravity conjecture. 
 
We will show that with a charged scalar field added (having sufficient charge) the previous Einstein-Maxwell solutions do become unstable. We also numerically construct the static solutions with scalar hair that they presumably settle down to and  study their properties.  We find  that the answer to (1) is no: with the scalar field present, the electric field never becomes arbitrarily large and the curvature remains bounded. This implies that one can no longer construct a counterexample to  cosmic censorship. Surprisingly, we also find that the minimum charge required to preserve cosmic censorship appears to agree precisely with the weak gravity bound.\footnote{Since our work is numerical, we cannot establish this rigorously. But the numerical data strongly suggests this is the case.} This shows a close connection between these two seemingly unrelated conjectures. At the moment, this connection seems rather mysterious and deserves further investigation.

However it seems clear that the weak gravity conjecture will not always preserve cosmic censorship. Another class of potential counterexamples in four dimensions with asymptotically AdS boundary conditions has been proposed which does not involve a Maxwell field. It is based on the superradiant instability of Kerr-AdS black holes \cite{Dias:2015rxy,Niehoff:2015oga} and would appear not to be affected by adding a charged scalar field.  We note however, that the violation of cosmic censorship likely to be exhibited by such classes of counterexamples is very different from the ones proposed in \cite{Horowitz:2016ezu} and observed in \cite{Crisford:2017zpi}. In particular, the violations where the weak gravity conjecture seems to play a role exhibit large curvatures in large regions of spacetime, whereas the ones with no Maxwell field are likely to lead to arbitrarily large curvatures in an arbitrarily small region of spacetime.
%%%%%%%%%%%%%%%%%%%%%%%%%%%%%%%%%%%%%%%%%%%
\section{Review of counterexamples to cosmic censorship}
Consider solutions to the bulk action
\begin{equation}
S = \frac{1}{16\pi G}\int \mathrm{d}^4 x\,\sqrt{-g}\left[R+\frac{6}{L^2}-F^{ab}F_{ab}\right]\,,
\label{eq:action0}
\end{equation}
where $L$ is the AdS length scale, and $F\equiv \dd A$ is the Maxwell field strength. With AdS boundary conditions, one is free to specify the (conformal) boundary metric at asymptotic infinity, as well as the asymptotic form of the vector potential $A_a$. We choose the boundary metric to be flat (as in the standard Poincar\'e coordinates for AdS)  
\begin{equation}
\dd s^2_\partial = -\dd t^2+\dd r^2+r^2 \dd\phi^2\;,
\label{eq:bndisflat}
\end{equation}
and the potential to asymptotically have only a nonzero time component
\begin{equation}
A|_\partial =a(t)\,p(r)\mathrm{d}t
\label{eq:bndgaugefield}
\end{equation}
where $a(t)$ is an amplitude and $p(r)$ is a radial profile that vanishes at large radius faster than $1/r$.  When $a$ is constant, static zero temperature solutions were found in  \cite{Horowitz:2014gva}. It was shown that these solutions all have a standard Poincar\'e horizon in the interior\footnote{When $p(r)$ falls off like $1/r$ or slower, this is no longer the case.}. One family of such solutions describe static, self-gravitating electric fields in AdS.  This family extends from $a=0$, where it meets with pure Poincar\'e AdS, to a maximum amplitude $a=a_{\max}$, where a naked curvature singularity appears. In \cite{Horowitz:2016ezu}, it was shown that this singularity extends for all $a> a_{\max}$.

The proposed counterexample to cosmic censorship \cite{Horowitz:2016ezu}, involves the following dynamical scenario.  Suppose $a(t)$ is initially zero, and slowly increases to a constant value larger than $a_{\max}$. If the amplitude is increased sufficiently slowly, then the bulk solution is well-approximated by a slowly evolving family of static solutions. If the endpoint of such an evolution is the singular static solution, then cosmic censorship will be violated. In  \cite{Crisford:2017zpi}, the full time dependent solution was found numerically (for the case where $p(r)$ falls off like $1/r$) and it was shown that $F^2$ does indeed grow as a power of time. This produces increasing curvature not just near the axis of symmetry, but everywhere along the horizon. Interestingly enough, the intrinsic geometry of the horizon  does not become singular. It is derivatives off the horizon that became large. 

We emphasize that unlike previous examples of naked singularities forming in four-dimensional spherical collapse scenarios \cite{Choptuik:1992jv,Bizon:2011gg}, our proposal is generic.  The function $a(t)$ does not need to be finely tuned. However, unlike those earlier examples, we do not expect that a naked singularity will form in finite time, only that the curvature will grow without bound.

Another branch of static solutions found in \cite{Horowitz:2014gva} describe hovering black holes. These are extremal spherical charged black holes which remain static since the normal gravitational attraction toward the Poincar\'e horizon is balanced by an electrostatic force toward infinity. Hovering black holes can only exist if $a$ is larger than some critical value, but when they are present, there is no violation of cosmic censorship: the solutions remain nonsingular as $a$ is increased and the hovering black hole just becomes larger. This branch of solutions does not affect the above counterexample to cosmic censorship since one starts with AdS initially and under evolution one can never form an extremal black hole since there is no charged matter. This will change when we add the charged scalar field and 
may provide another way to save cosmic censorship.

%%%%%%%%%%%%%%%%%%%%%%%%%%%%%%%%%%%%%%%%%%%
\section{The weak gravity conjecture in AdS}
Before proceeding, we will attempt to sharpen the weak gravity conjecture in AdS. The weak gravity conjecture in essence ensures that   extremal charged black holes are quantum mechanically unstable to Schwinger pair production. In flat space, this is a purely quantum mechanical instability, but in AdS it turns out to give rise to a classical instability. This instability is the so called charged superradiant instability \cite{Starobinskil:1974nkd,Gibbons:1975kk} whose endpoint has been studied over the past decade \cite{Basu:2010uz,Bhattacharyya:2010yg,Dias:2011tj,Gentle:2011kv,Markeviciute:2016ivy,Bosch:2016vcp,Dias:2016pma}.

In AdS, the onset of this superradiant instability depends on the size of the black hole. To stay as close as possible to the original weak gravity conjecture, we will consider an arbitrarily small black hole.  The general spherically symmetric  charged black hole is the Reissner-Nordstr\"om (RN) black hole in AdS, which is a solution of the equations of motion derived from (\ref{eq:action0}) and takes the following simple form:
\begin{equation}
\mathrm{d}s^2=-f(r)\mathrm{d}\hat{t}^2+\frac{\mathrm{d}r^2}{f(r)}+r^2\mathrm{d}\Omega^2\,,\quad A = \mu\,\left(1-\frac{r_+}{r}\right)\mathrm{d}\hat{t}
\end{equation}
where $\mathrm{d}\Omega^2$ is the metric on a unit round 2-sphere and
\begin{equation}
f(r)=\frac{r^2}{L^2}+1+\frac{\mu^2\,r_+^2}{r^2}-\frac{r_+}{r}\left(\frac{r_+^2}{L^2}+1+\mu^2\right)\,.
\end{equation}
The event horizon is a null hypersurface with $r=r_+$, where $r_+$ is the largest positive real root of $f(r)$. One can show that the Hawking temperature of a RN black hole in AdS is given by
\begin{equation}
T_H=\frac{L^2-L^2 \mu ^2+3 r_+^2}{4 L^2 \pi  r_+}\,.
\end{equation}
The event horizon becomes degenerate when $T_H=0$, which occurs for $\mu=\mu_{\mathrm{ext}}\equiv\sqrt{1+3r_+^2/L^2}$. 

Superradiant scattering for a scalar field of mass $m$ and charge $q$ occurs if \cite{Starobinskil:1974nkd,Gibbons:1975kk}
\begin{equation}
0<\hat{\omega}<q\,\mu\,,
\label{eq:super}
\end{equation}
where $\hat{\omega}$ is the frequency of the perturbation we are considering. So we need to know what $\hat{\omega}$ are allowed for a small black hole, \emph{i.e.} what the quasinormal mode spectrum of a small extremal RN black hole looks like. If the RN black holes are small, two decoupled sectors of quasinormal mode excitations exist: one whose imaginary part grows infinitely negative as the size of the hole decreases, and another whose imaginary part drops to zero as the black hole becomes smaller and whose real part approaches the normal modes of AdS \cite{Berti:2003ud,Uchikata:2011zz}. It is the latter type that is of interest to us. In the approximation where the extreme black hole is very small, the quasinormal mode with the smallest real part will have $\hat{\omega}\,L= \Delta+\smallO(r_+/L)$ where
\begin{equation}\label{eq:Delta}
\Delta = \frac{3}{2}+\sqrt{\frac{9}{4}+L^2\,m^2}\,.
\end{equation}
Substituting in Eq.~(\ref{eq:super}), and noting that small extremal black holes have $\mu=1+\mathcal{O}(r_+^2/L^2)$, gives the following lower bound on the scalar field charge $q$
\begin{equation}
q \ge q_{W}\equiv \frac{\Delta}{L}\,.
\end{equation}
This is the bound that might prevent the violation of the weak cosmic censorship presented in the previous section. Note that this is essentially the complement of the BPS bound in AdS \cite{Denef:2009tp}.  
%%%%%%%%%%%%%%%%%%%%%%%%%%%%%%%%%%%%%%%%%%%
\section{Adding charge carriers}
To satisfy the weak gravity conjecture, we now augment (\ref{eq:action0}) by adding a minimally coupled charged scalar field $\Phi$. The new action reads
\begin{equation}
S = \frac{1}{16\pi G}\int \mathrm{d}^4 x\,\sqrt{-g}\left[R+\frac{6}{L^2}-F^{ab}F_{ab}-4\,(\mathcal{D}_a \Phi)(\mathcal{D}^a \Phi)^\dagger-4\,m^2\Phi \Phi^\dagger\right]\,,
\label{eq:action}
\end{equation}
where $\mathcal{D}_a = \nabla_a-i\,q\, A_a$ is the gauge covariant derivative with respect to $A_a$, $m$ is the charged scalar field mass and $q$ its charge. The following equations of motion can be derived from (\ref{eq:action}):
\begin{subequations}
\begin{align}
&R_{ab} +\frac{3}{L^2}g_{ab}= 2\left(F_{a}^{\phantom{a}c}F_{bc}-\frac{g_{ab}}{4}F_{cd}F^{cd}\right)+2(\mathcal{D}_a \Phi) (\mathcal{D}_b \Phi)^\dagger+2(\mathcal{D}_a \Phi)^\dagger (\mathcal{D}_b \Phi)+2\,m^2 g_{ab}\Phi\Phi^\dagger\,,\label{eq:einstein}
\\
&\nabla_a F^{a}_{\phantom{a}b}=i\,q\,\left[(\mathcal{D}_b \Phi)\Phi^\dagger-(\mathcal{D}_b \Phi)^\dagger\Phi\right]\,,
\\
&\mathcal{D}_a\mathcal{D}^a \Phi = m^2 \Phi\,.
\label{eq:scalar}
\end{align}
\label{eqs:motion}
\end{subequations}

In order to find novel static solutions of (\ref{eqs:motion}) we are going to use the DeTurck method. This method was first introduced in \cite{Headrick:2009pv}, and was recently reviewed in great detail in \cite{Dias:2015nua}. We deform (\ref{eq:einstein}) and consider instead:
\begin{equation}
R_{ab}+\frac{3}{L^2}g_{ab}-\nabla_{(a}\xi_{b)} =2\left(F_{a}^{\phantom{a}c}F_{bc}-\frac{g_{ab}}{4}F_{cd}F^{cd}\right)+2(\mathcal{D}_a \Phi) (\mathcal{D}_b \Phi)^\dagger+2(\mathcal{D}_a \Phi)^\dagger (\mathcal{D}_b \Phi)+2\,m^2 g_{ab}\Phi\Phi^\dagger\,,\label{eq:deturck}
\end{equation}
where $\xi^a = \left[\Gamma^a_{cd}(g)-\Gamma^a_{cd}(\bar{g})\right]g^{cd}$ and $\Gamma^{a}_{bc}(\mathfrak{g})$ is the Levi-Civitta connection associated with a metric $\mathfrak{g}$. $\bar{g}$ is a reference metric which, as we shall see, controls our gauge choice. In order for this method to work, $\bar{g}$ must have the same asymptotic structure, including horizon location and asymptotic infinity, as the metric we wish to find.

The advantage of solving (\ref{eq:deturck}) instead of (\ref{eq:einstein}) is tremendous: if we are interested in finding static solutions, (\ref{eq:deturck}) comprises a system of elliptic equations which can be readily relaxed using a Newton-like method. One might worry whether solutions of (\ref{eq:deturck}) will necessarily be solutions of (\ref{eq:einstein}), that is to say, whether solutions of (\ref{eq:deturck}) necessarily have $\xi^a=0$. Under certain circumstances one can show that solutions of Eq.~(\ref{eq:deturck}) with $\xi^a\neq0$ cannot exist \cite{Figueras:2011va,Figueras:2016nmo}. However, this proof does not extend easily if matter fields are included. The idea is to solve (\ref{eq:deturck}) and check \emph{a posteriori} if $\chi \equiv \xi^a\xi_a$ approaches zero in the continuum limit\footnote{Note that since we are interested in static solutions, one can show that $\chi$ is necessarily positive, so we do not need to check $\xi^a$ component by component.}. Since the equations are elliptic, we know that solutions are locally unique, therefore solutions with $\chi\neq0$ cannot be arbitrarily close to solutions with $\chi=0$.

The reason why (\ref{eq:einstein}) does not give rise to a well defined elliptic problem has to do with the fact that general relativity is coordinate invariant. As such, a particular gauge has to be chosen before a given problem is solved. From the condition $\xi^a=0$, one can see that the DeTurck method is simply a rewritting of the original Einstein equation (\ref{eq:einstein}) in generalised harmonic coordinates $\triangle x^ a = \Gamma^a_{cd}(\bar{g})g^{cd}$. This in turn implies that the choice of $\bar{g}$ is ultimately connected with a choice of gauge.

\subsection{Review of Einstein-Maxwell solutions\label{subsec:uncharged}}
For the moment we will focus on solutions with $\Phi=0$, and reconstruct the solutions presented in \cite{Horowitz:2014gva}. We are going to use the coordinate system constructed in \cite{Horowitz:2014gva}, which is well adapted to zero temperature horizons. We consider solutions whose conformal boundary metric approaches Minkowski space (\ref{eq:bndisflat}). Furthermore, we are interested in axisymmetric and static solutions, so we adapt our coordinate system such that $\partial_t$ and $\partial_\phi$ are Killing fields.

To motivate the coordinate system used in \cite{Horowitz:2014gva}, we start with the metric corresponding to pure AdS in Poincar\'e coordinates. Such line element reads
\begin{equation}
\mathrm{d}s^2=\frac{L^2}{z^2}\left[-\dd t^2+\dd r^2+r^2 \dd\phi^2+\dd z^2\right]\,.
\label{eq:pure}
\end{equation}
We  regard the $(r,z)$ coordinates as Cartesian coordinates, and introduce polar-like coordinates in the following way:
\begin{subequations}
\begin{align}
&z=\frac{y \sqrt{2-y^2}}{1-y^2}(1-x^2)\,,
\\
&r=\frac{y \sqrt{2-y^2}}{1-y^2}x\sqrt{2-x^2}\,.
\end{align}
\end{subequations}
The line element (\ref{eq:pure}) written in $(x,y)$ coordinates becomes:
\begin{equation}
\dd s^2=\frac{L^2}{\left(1-x^2\right)^2}\left[-\frac{\left(1-y^2\right)^2\dd t^2}{y^2 \left(2-y^2\right)}+\frac{4\,\dd x^2}{2-x^2}+\frac{4\,\dd y^2}{y^2 \left(1-y^2\right)^2\left(2-y^2\right)^2}+x^2 \left(2-x^2\right)\dd\phi^2\right]\,.
\label{eq:pureads}
\end{equation}
$x$ is like an angular coordinate and $y$ is like a radial coordinate. Both take values in $[0,1]$. The Poincar\'e horizon is now located at $y=1$, and the axis of rotation is located at $x=0$. The conformal boundary is located at $x=1$, and $y=0$ denotes the intersection of the conformal boundary with the axis of symmetry.

We want to consider a gauge potential that asymptotically, \emph{i.e.} when approaching $x=1$, has only a nonzero time component:
\begin{equation}
A_\partial = \frac{a\,\dd t}{\displaystyle \left(1+\frac{r^2}{\ell^2}\right)^{\frac{n}{2}}}\,.
\label{eq:profile}
\end{equation}
Due to the underlying conformal invariance of the theory deep in the UV, only the product $a\,\ell$ is physically meaningful. From here onwards we will set $\ell=1$ without loss of generality. In terms of the $y$ coordinates the asymptotic profiles (\ref{eq:profile}) read
\be
A_\partial = a\,(1-y^2)^{n}\dd t\,.
\ee

We now wish to find the family of static solutions with increasing $a$. Before presenting our results, let us mention that when using the DeTurck method, one has to write down the most general \emph{ansatz} compatible with the symmetries of our problem. Given that we are interested in solutions that are static and axisymmetric, this restricts our \emph{ansatz} to take the following form
\begin{subequations}
\begin{multline}
\dd s^2=\frac{L^2}{\left(1-x^2\right)^2}\Bigg[-\frac{\left(1-y^2\right)^2\,Q_1\,\dd t^2}{y^2 \left(2-y^2\right)}+\frac{4\,Q_4}{2-x^2}\left(\dd x+\frac{Q_3}{1-y^2}\dd y \right)^2\\
+\frac{4\,Q_2\,\dd y^2}{y^2 \left(1-y^2\right)^2\left(2-y^2\right)^2}+x^2 \left(2-x^2\right)\,Q_5\,\dd\phi^2\Bigg]\,,
\end{multline}
\hspace{0.1cm}
\begin{equation}
A= L\, Q_6\,\dd t\,,
\end{equation}
\label{eqs:ansatz}
\end{subequations}
where the $Q_i$, with $i\in\{1,\ldots,6\}$, depend on $x$ and $y$ and are the functions we intend to determine. For the reference metric we take pure AdS written in $x$ and $y$ coordinates, that is to say, we take the line element (\ref{eq:pureads}).

Finally, we discuss the issue of boundary conditions. At the conformal boundary, located at $x=1$, we impose
\be
Q_1=Q_2=Q_4=Q_5=1\,,\quad Q_3=0\,, \quad \text{and}\quad Q_6 = a\,(1-y^2)^{n}\,.
\ee
At the symmetry axis, located at $x=0$, we find
\be
\frac{\partial Q_1}{\partial x}=\frac{\partial Q_2}{\partial x}=\frac{\partial Q_4}{\partial x}=\frac{\partial Q_5}{\partial x}=\frac{\partial Q_6}{\partial x}=0\,, \quad Q_4=Q_5\,, \quad \text{and}\quad Q_3=0\,.
\ee
At the point $y=0$ (corresponding in the original coordinates of (\ref{eq:pure}) to $r=z=0$) we demand
\be
Q_1=Q_2=Q_4=Q_5=1\,,\quad Q_3=0\,, \quad \text{and}\quad Q_6 = a\,.
\ee
In \cite{Horowitz:2014gva}, it was observed that the IR, for $n>1$,  always has a Poincar\'e horizon. This corresponds to setting at $y=1$
\be
Q_1=Q_2=Q_4=Q_5=1\,,\quad Q_3=0\,, \quad \text{and}\quad Q_6=0\,.
\ee

Our results are completely consistent with those of \cite{Horowitz:2014gva}, and are described there. In particular, we find that solutions with a single connected horizon exist up to a maximum amplitude $a_{\max}$. This maximum amplitude strongly depends on $n$ (see Table \ref{tab:1}).
\begin{table}
\centering
\begin{tabular}{|c|c|c|c|c|c|}
\hline
$n$ & 2 &   4   & 6 & 8 & 10
                     \\
                     \hline
$a_{\max}$   & 2.64 & 4.88 & 6.48 & 7.97 & 9.02
\\
\hline
\end{tabular}
\caption{Maximum amplitude for several values of $n$.}
\label{tab:1}
\end{table}
%%%%%%%%%%%zero-mode%%%%%%%%%%%%%%%%%%%%%%%%%%%%%%%%
\subsection{Stability}

We now wish to test whether the  solutions we have constructed are unstable to forming a scalar field condensate using the mechanism proposed in \cite{Gubser:2008px,Hartnoll:2008vx}. In order to do this, we will first search for zero-modes following the same strategy as in \cite{Horowitz:2013jaa,Costa:2017tug}. A static normalizable mode usually marks the transition between stable and unstable solutions. We thus consider  Eq.~(\ref{eq:scalar}) on the fixed  backgrounds constructed in section \ref{subsec:uncharged}. Time independent solutions of (\ref{eq:scalar}) on a fixed  background reduces to solving the following linear partial differential equation
\begin{equation}
(\nabla_a \nabla^a -m^2)\Phi = q^2\,A_a A^a\,\Phi\,.
\label{eq:eige}
\end{equation}
This equation takes the form of a generalised eigenvalue problem, with eigenfunction $\Phi$ and eigenvalue $q^2$. For each profile constructed in the previous section and for each value of $m$, we can determine the smallest eigenvalue which we will denote $q_{\min}^2$. 

Our numerical method only allows us to study scalars with $\Delta\geq1$. 
When $\Delta$ is small, one has a choice of boundary conditions for $\Phi$. We will choose boundary conditions so that $\Phi$ is the holographic dual of a boundary operator with conformal dimension $\Delta\geq1$ \eqref{eq:Delta}. This means keeping the slower fall-off behavior for $1\leq\Delta<3/2$ (sometimes called ``alternative quantization") and requiring the faster fall-off for everything else. For comparison with the predictions from the weak gravity conjecture, we will parametrize the dependence on the mass using $\Delta$ instead of $m$.

\begin{figure}
\includegraphics[width=.57\textwidth]{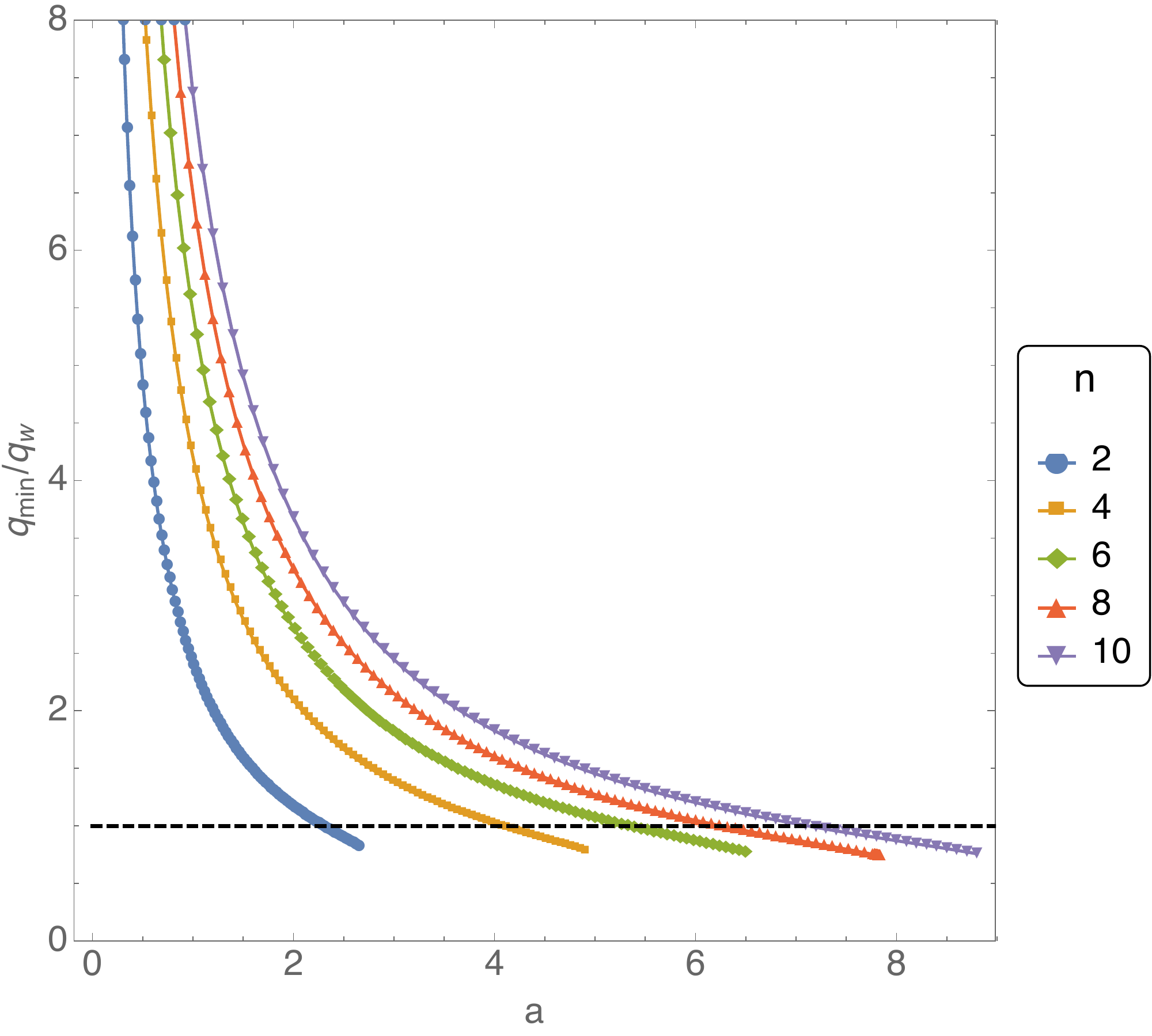}
\caption{The minimum charge, $q_{\min}$, needed for a zero-mode as a function of the amplitude $a$, plotted for several different profiles. From bottom to top we have $n=2\,,4\,,6\,,8\,,10$, respectively. The horizontal dashed line represents the weak gravity bound $q_{\min}/q_W=1$. These curves were determined for $\Delta=2$.}
\label{fig:delta2severaln}
\end{figure}

\begin{figure}
\includegraphics[width=.6\textwidth]{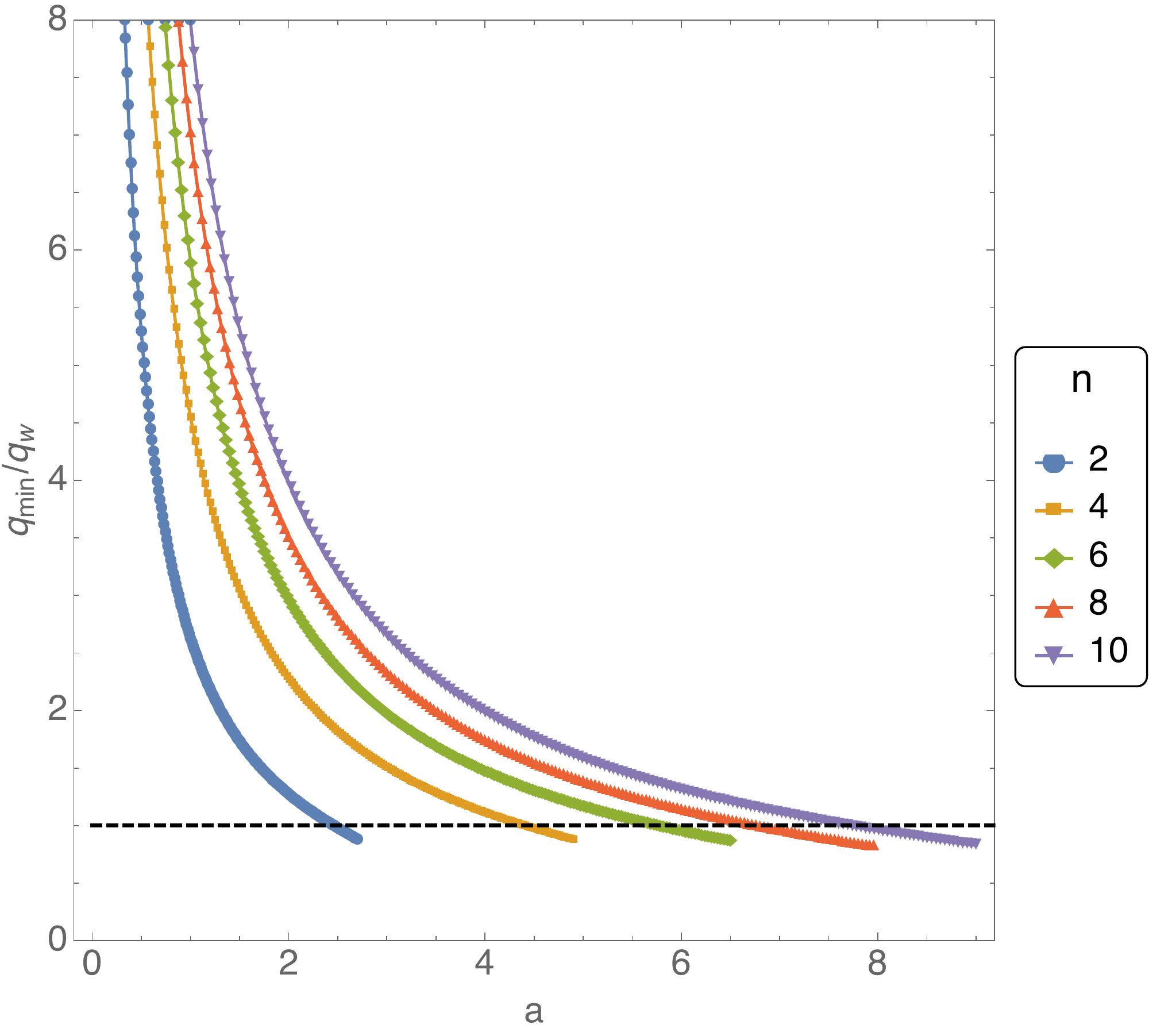}
\caption{Similar to Fig. 1, but now for $\Delta=4$.}
\label{fig:qmin_delta_4}
\end{figure}

To begin, we will take $\Delta = 2$ and compute $q_{\min}$ for several different profiles $p(r)$ \eqref{eq:profile} labelled by $n$, and several different amplitudes $a$. The results are displayed in Fig.~(\ref{fig:delta2severaln}). As expected, for each  profile $q_{\min}$ is a decreasing function of $a$ because as we increase the amplitude, we increase the electric field in the bulk which should  make it easier to trigger the scalar instability.  To facilitate our later comparison with the weak gravity bound, we plot $q_{\min}/q_W$ on the vertical axis, and plot a horizontal line at $q_{\min}/q_W = 1$. 
These curves do not change qualitatively if we change $\Delta$. For example, Fig.~(\ref{fig:qmin_delta_4}) shows the results for $\Delta=4$.

\begin{figure}
\includegraphics[width=1\textwidth]{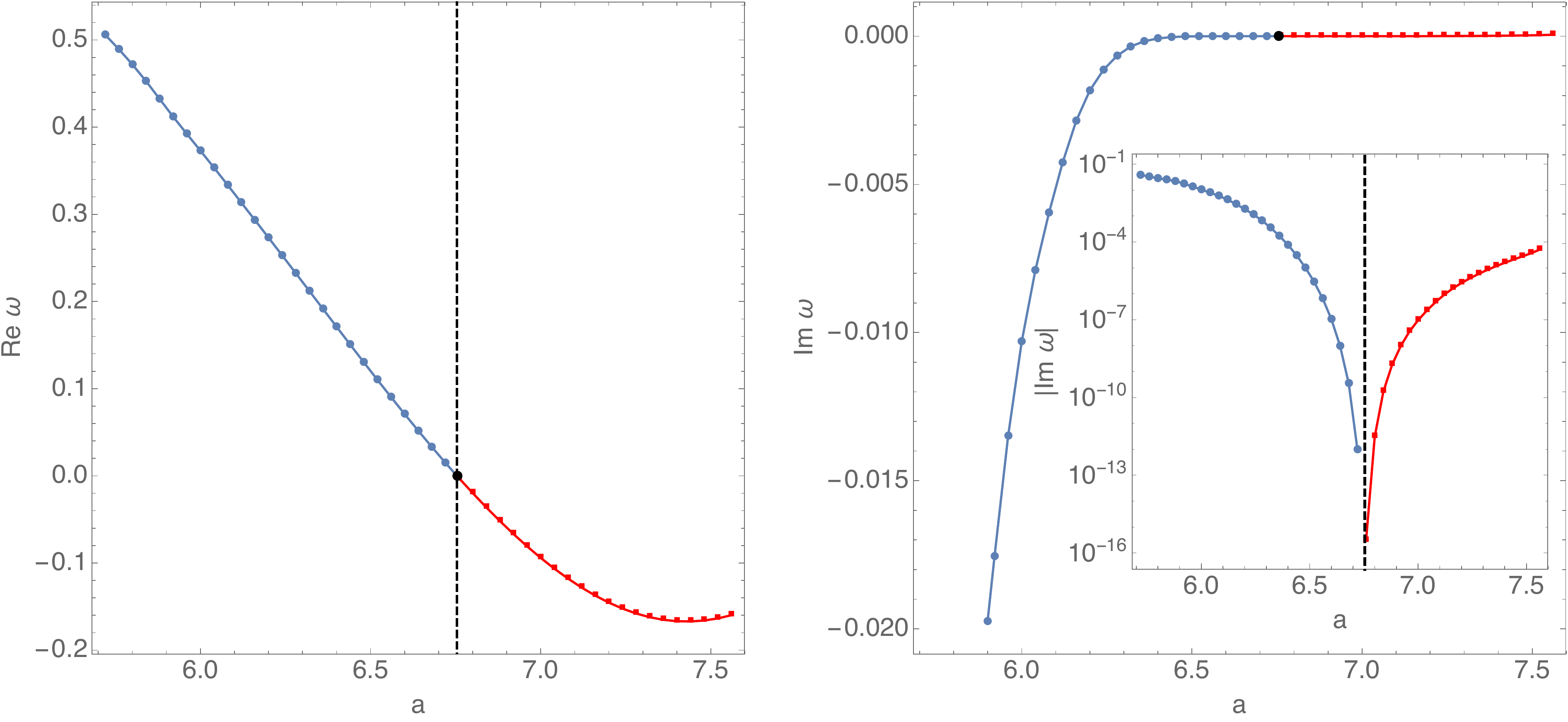}
\caption{The lowest quasinormal mode frequency for $n=8$, $\Delta = 4$, and $q=q_W$.  The black dot denotes the zero-mode computed directly from Eq.~(\ref{eq:eige}). The red dots have the opposite sign of $\omega$ from the blue dots. The insert on the right plots the data on a logarithmic scale, clearly showing that  $\mathrm{Im} \ \omega $ becomes  positive after the zero-mode, so the solution without the scalar field becomes unstable.}
\label{fig:instability}
\end{figure}

We next want to check whether these zero-modes  indeed mark  the boundary between stable and unstable solutions.  To do this  we  include harmonic time dependence $e^{-i\omega t}$ in the scalar perturbation and compute the lowest quasinormal mode frequency for each $a$. We 
take $\Delta = 4$, and set  $q=q_W$ since larger $q$ are more likely to induce instabilities.  Since the zero-mode has $\omega =0$, as we change $a$ at fixed charge, both the real and imaginary parts of the lowest quasinormal mode frequency must pass through zero. If $\mathrm{Im} \ \omega $ becomes  positive, the mode becomes unstable. As Fig.~(\ref{fig:instability}) shows, this is exactly what we find. The location of the zero-mode is shown as a black dot, and $\mathrm{Im} \ \omega $ changes sign there. Since $\mathrm{Im} \ \omega $ is small near the  zero-mode, the instability sets in slowly. We believe that the instability will set in faster for larger values of $q$.

Returning to Figs.~(\ref{fig:delta2severaln}) and (\ref{fig:qmin_delta_4}), this shows that for each profile, the original Einstein-Maxwell solutions  are stable below the curve, but unstable above it. The curves end at $a_{\max}$ and the solutions (with $\Phi =0$) are singular for larger $a$. The fact that all the curves in Figs.~(\ref{fig:delta2severaln}) and (\ref{fig:qmin_delta_4}) cross the line $q_{\min}/q_W = 1$ means that if we assume the weak gravity conjecture, the  counterexamples to cosmic censorship proposed in \cite{Horowitz:2016ezu,Crisford:2017zpi} must be reexamined. If we start with any $q>q_W$, and slowly increase the amplitude $a$, the Einstein-Maxwell solution becomes unstable before it becomes singular. So the previous singular solution is no longer applicable, and one must study the solutions with nonzero scalar field included.

\begin{figure}
\includegraphics[width=.47\textwidth]{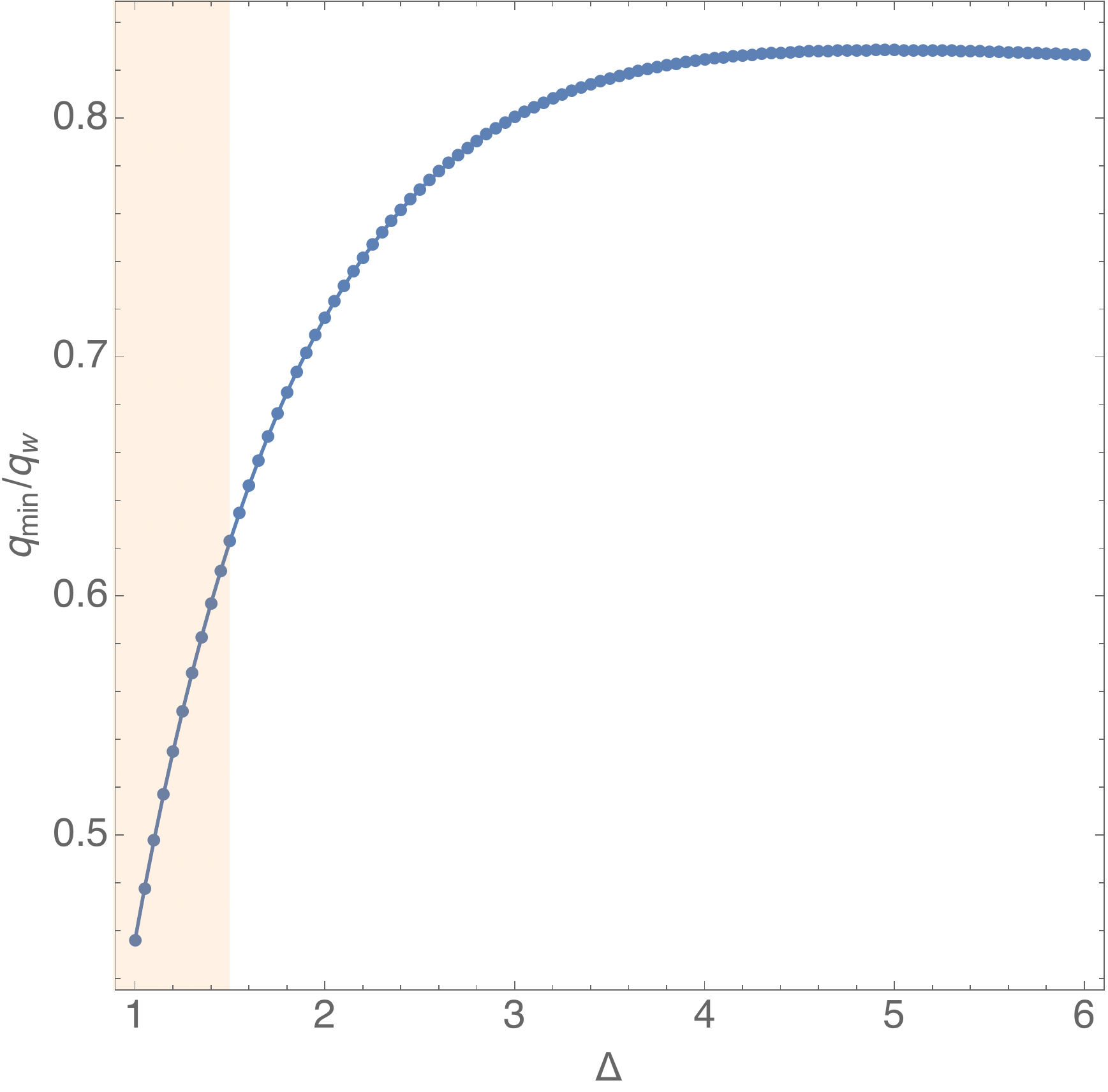}
\caption{$q_{\min}/q_W$ as a function of $\Delta\geq1$ plotted for $n=8$ and $a=a_{\max}$. The orange region indicates the region of moduli space where we used alternative 
boundary conditions.}
\label{fig:qminfixdelta}
\end{figure}

Before we do so, we  first  ask if 
this holds for all values of the mass of the scalar field. To answer this, we repeated the  calculation of the zero-mode charge $q_{\min}$ for many values of $\Delta$. Since  the lowest value of $q_{\min}$ always occurs for $a=a_{\max}$, we keep $a=a_{\max}$ and compute $q_{\min}$ for various $\Delta$ for the $n=8$ profile. The results are plotted in Fig.~(\ref{fig:qminfixdelta}). The fact that  $q_{\min}/q_W$ is always less than one means that the situation is always qualitatively the same as the 
previous cases. If we assume the weak gravity conjecture, the previous counterexamples to cosmic censorship are no longer valid and we must study solutions with nonzero scalar field to see what happens. This is what we turn to next.

%%%%%%%%%%%%%%%%%%%%%%%%%%%%%%%%%%%%%%%%%%%
\subsection{Nonlinear charged scalar hair}
In this section we will construct the nonlinear  solutions with scalar hair that branch from the zero-modes constructed previously. For concreteness we will present results for $n=8$ and $\Delta = 2$, but other values of $n$ and $\Delta$ have similar conclusions. The choice $\Delta=2$ is convenient for the particular numerical method we use in our integration scheme. Irrational values of $\Delta\geq1$ would produce an asymptotic decay close to the conformal boundary with irrational powers. This in turn would lead to weak convergence for spectral collocation methods. To bypass this, we will take $\Delta = 2$, corresponding to a scalar field with mass $m^2=-2/L^2$ and 
requiring the standard (faster fall-off) boundary conditions.\footnote{Other values of $\Delta$ are compatible with exponential convergence, for instance $2\Delta = \mathbb{Z}$. }

The metric and gauge field ans\"atze will remain as in Eqs.~(\ref{eqs:ansatz}), and for the scalar field we take
\begin{equation}
\Phi = \left(1-x^2\right)^2 y^2 \left(2-y^2\right)\,Q_7\,,
\end{equation}
where the powers of $x$ and $y$ were chosen to make $Q_7(y,1)$ directly proportional to the expectation value of the operator dual to $\Phi$, which we coin $\langle \mathcal{O}_2\rangle.$\footnote{It turns out that the precise relation is $\langle \mathcal{O}_2\rangle=(1- y^2)^2 Q_7(y,1)$.}

\begin{figure}
\includegraphics[width=.47\textwidth]{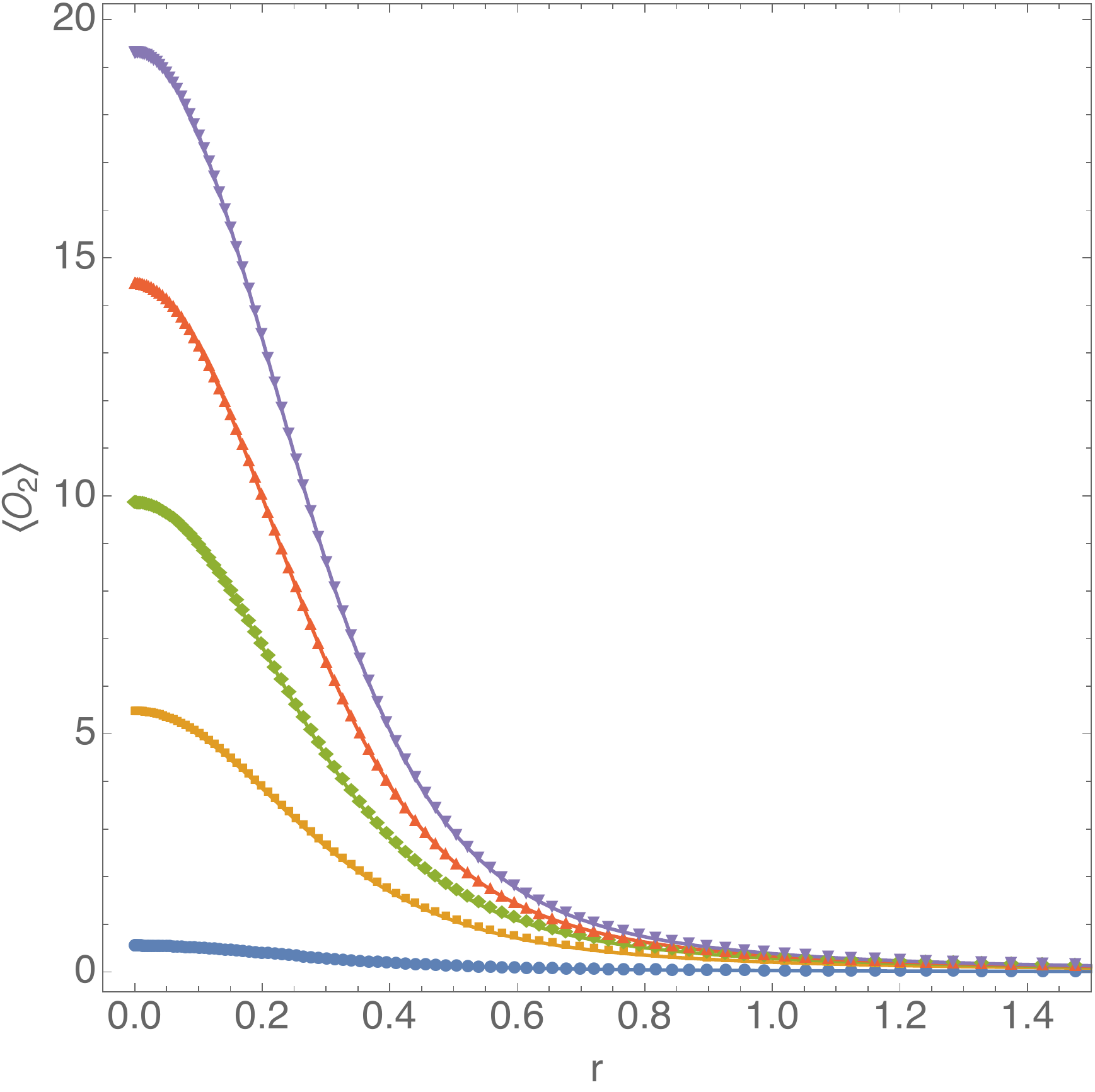}
\caption{$\langle \mathcal{O}_2 \rangle$ as a function of $r$, for several values of $a$ and with $n=8$; from top to bottom we have $a= 10\,, 
9.0\,, 8.0\,, 7.0\,,6.27$.}
\label{fig:scalarnonlinear}
\end{figure}

\begin{figure}
\includegraphics[width=.47\textwidth]{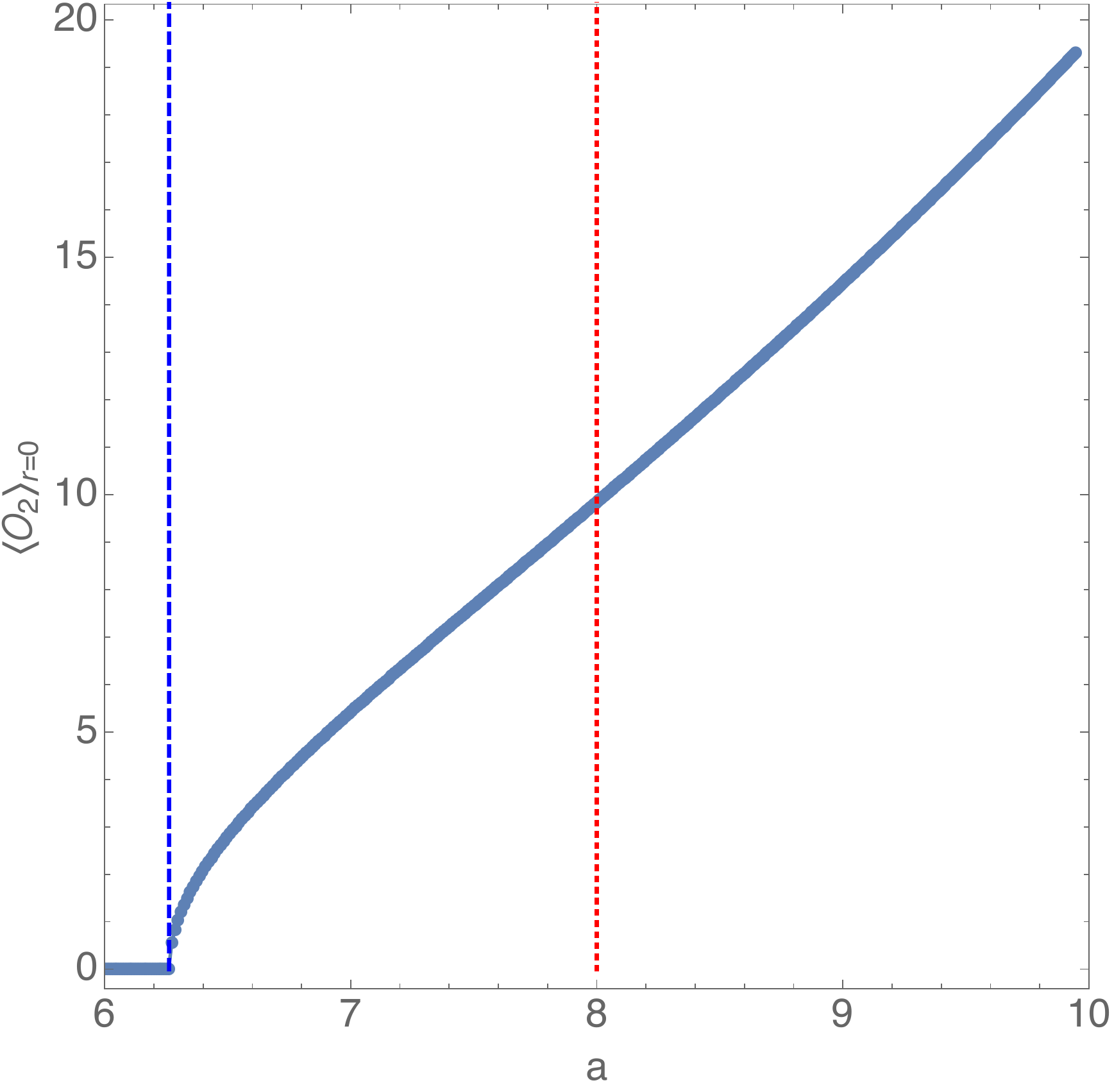}
\caption{$ \langle \mathcal{O}_2\rangle$ at $r=0$ plotted as a function of $a$ for profile $n=8$, with the vertical dashed blue line representing the onset computed in the previous section and the vertical red dotted line corresponding to $a=a_{\max}$.}
\label{fig:maxO}
\end{figure}

To preserve cosmic censorship, we must  show that if $q\geq q_W$ and $a$ is increased, a smooth scalar field condenses and exists for arbitrarily large values $a$. We will take $q=q_W$, since that is the most difficult case to condense. We find that solutions with $\Phi \ne 0$ indeed exist for all amplitudes larger than the zero-mode shown in Fig. 1.  In Fig.~(\ref{fig:scalarnonlinear}) we show the  expectation value for the operator dual to $\Phi$ as a function of the boundary radial coordinate $r$, for several values of $a$.
The scalar condensate is clearly largest at the origin, $r=0$, and falls off for larger $r$. These solutions can thus be viewed as localized holographic superconductors. Increasing the amplitude increases the maximum of the condensate. To see this more clearly, 
in Fig.~(\ref{fig:maxO}) we present $\langle \mathcal{O}_2\rangle$ at $r=0$ as a function of $a$, and to help guide the eye we also show two vertical lines, corresponding to the onset of the scalar condensate deduced in the previous section (blue dashed line) and  $a=a_{\max}$ (red dotted line), both for $n=8$. Recall that when $\Phi = 0$, $a_{\max}$ is the amplitude at which the solution becomes singular. Clearly, solutions with scalar hair exist for much larger amplitudes and seem to exist for arbitrarily large values of $a$! Thus if we assume the weak gravity conjecture, there will be a field with $q=q_W$ and one cannot violate cosmic censorship by slowly increasing the amplitude on the boundary by any finite amount. For any final value of $a$, there is a static nonsingular  bulk solution for the geometry to settle down to.

 To investigate the connection between cosmic censorship and the weak gravity conjecture  more quantitatively, we now ask if we could lower $q$ below $q_W$ and still preserve cosmic censorship. From Figs.~(\ref{fig:delta2severaln}),  (\ref{fig:qmin_delta_4}) and (\ref{fig:qminfixdelta}) it would appear that the answer is yes: one can indeed lower the charge slightly below the weak gravity bound and still have the scalar field condense before reaching a singularity. However, this does not take into account the possibility that naked singularities can still form even with the scalar field nonzero.

\begin{figure}
\includegraphics[width=.47\textwidth]{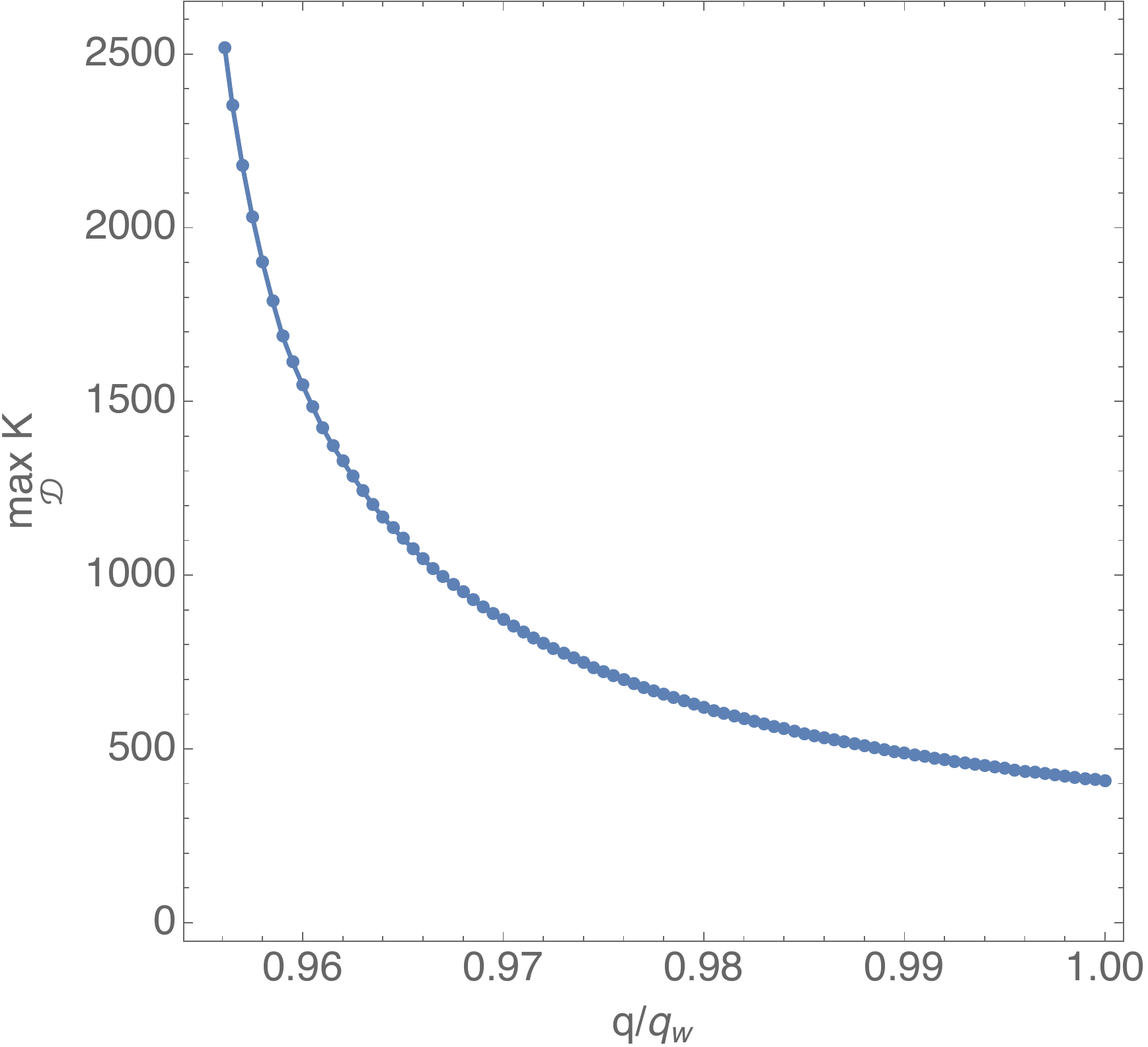}
\caption{Maximum of Kretschmann scalar over our integration domain as a function of $q/q_W$ for fixed $a=8.5>a_{\max}$ with $n=8$ and $\Delta = 2$.}
\label{fig:kret}
\end{figure}

To explore this possibility, we ask what happens to the full nonlinear solutions with $\Phi\ne 0$ if we lower $q$ keeping $a>a_{\max}$.  We take for instance a solution with $n=8$, at fixed $a=8.5>a_{\max}$, and lower $q$ as much as we can, while monitoring a curvature invariant such as the Kretschmann scalar, defined here as $K\equiv L^4R_{abcd}R^{abcd}$. Our findings are plotted in Fig.~(\ref{fig:kret}) where we see that the maximum of Kretschmann scalar outside the Poincar\'e horizon seems to blow up at some critical value of $q=q_c<q_W$. In order to find the precise value at which $\underset{\mathcal{D}}{\max} K$ diverges, we monitor $(\underset{\mathcal{D}}{\max} K)^{-1}$ and use linear extrapolation. 

\begin{figure}
\includegraphics[width=1\textwidth]{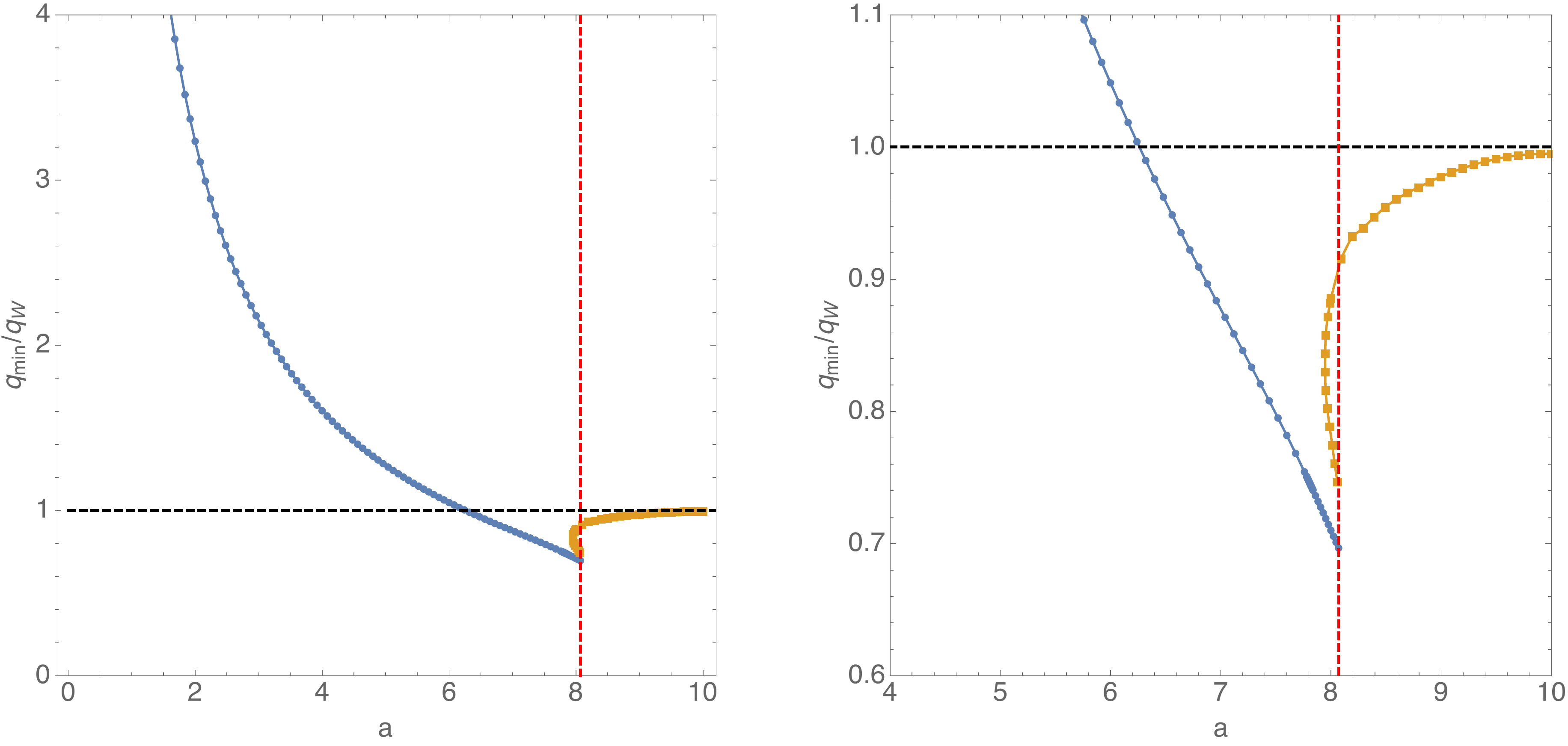}
\caption{Phase diagram of solutions for $n=8$ and $\Delta =2$. The dashed vertical line denotes $a=a_{\max}$. Solutions with $\Phi \ne 0$ exist above the (blue and yellow) line connecting the dots,  and $\Phi \to 0$  along this line for $a< a_{\max}$, but develops singularities along the line for  $a> a_{\max}$. The right panel shows a blow-up of part of the left panel.}
\label{fig:last}
\end{figure}

One can repeat this exercise for several values of $a$.  Remarkably, the singularity in the solutions with $\Phi \ne 0$ appears to approach the weak gravity bound as $a$ increases! This is shown in Fig.~(\ref{fig:last}),\footnote{We thank Yong-Qiang Wang for pointing out an error in an earlier version of this plot.} where we have added the minimum charge for the $\Phi \ne 0$ solutions with $a>a_{\max}$ to our earlier plot of the minimum charge for $a<a_{\max}$. It is difficult to push the numerics to larger values of the amplitude, but at $a = 10.0$ the singularity appears when the charge is
just half a percent lower than $q_W$ and this difference is clearly decreasing with $a$.   This means that if we take $q$ below $q_W$ and slowly increase the amplitude, even though our previous Einstein-Maxwell solutions become unstable and the scalar field turns on, it will still become singular at a finite value of $a$. So one can still violate cosmic censorship this way. Thus the bound on the charge to preserve cosmic censorship appears to be precisely the weak gravity bound.

\section{Final comments}

Motivated by the weak gravity conjecture, we have added a charged scalar field to our earlier counterexamples to cosmic censorship in AdS. 
We have presented strong evidence that the weak gravity conjecture bound on the charge is both necessary and sufficient to preserve cosmic censorship for this class of examples. We find this connection to  be remarkable and clearly deserves further investigation. 

Perhaps one step toward understanding this connection is the following. In the solutions with $\Phi \ne 0$ and $a>a_{\max}$,  when we lower the charge and the curvature becomes large, the scalar field also becomes large at the same location. This produces a large localized charge density with a geometry perhaps similar to that outside a small charged black hole. But if the charge is below the weak gravity bound, a small charged black hole cannot have smooth scalar hair. The scalar field necessarily diverges at the horizon. This might help explain why we are obtaining the same bound on the charge.

As in \cite{Horowitz:2016ezu}, our arguments about the validity of cosmic censorship are based on the structure of the space of static solutions. This is reasonable since one can imagine slowly increasing the amplitude $a$ on the boundary, and  the bulk is expected to remain close to the static solutions (whenever they are nonsingular). However we have not yet done the complete time dependent evolution (analogous to \cite{Crisford:2017zpi}),  and there is value in doing so.
As mentioned in section II, the original Einstein-Maxwell theory has another branch of solutions describing hovering charged black holes. These do not affect the original proposed counterexample to cosmic censorship since they are not present initially and cannot form under evolution since there is no charged matter. Since we have now added a charged scalar field, one should ask if hovering black holes could form as we slowly increase the amplitude on the boundary. This might be possible for $q<q_W$ near the amplitude where the static solution becomes singular. If so, the weak gravity conjecture would still preserve cosmic censorship, but it might also be preserved for some $q< q_W$. One needs to do a complete time dependent evolution to see if this is the case.

Finally, as we mentioned earlier, the smooth solutions we have constructed with $\Phi \ne 0$ can be viewed as localized holographic superconductors. It might be interesting to study their critical temperatures and some of their transport properties. 

%----- Acknowledgements -------
\vskip 1cm
\centerline{\bf Acknowledgements}
GH was supported in part by NSF grant PHY-1504541. TC was supported by an STFC studentship. JES was supported in part by STFC grants PHY-1504541 and ST/P000681/1.
\vskip .5 cm
%%%%%%%%%%%%%%%%%%%%%%%%%%%%%%%%%%%%%%%%%%%

%%%%%%%%%%%%%%%%%%%%%%%%%%%%%%%%%%%%%%%%%%%

\bibliographystyle{JHEP}
\bibliography{all}
  
\end{document}